\documentstyle[preprint,eqsecnum,aps]{revtex}
\preprint{\vbox{
\hbox{IFT-P.005/95}
\hbox{December 1994}
\hbox{hep-ph/9412304}
}}
\begin{document}
\draft
\title{
New fermions and a vector-like third generation \\ in
$SU(3)_c\otimes SU(3)_L\otimes U(1)_N$ models}
\author{ Vicente Pleitez}
\address{
Instituto de F\'\i sica Te\'orica\\
Universidade Estadual Paulista\\
Rua Pamplona, 145\\
01405-900-- S\~ao Paulo, SP\\
Brazil}
\maketitle
\begin{abstract}
We study two 3-3-1 models with i) five (four) charge 2/3 ($-1/3$)
quarks and, ii)
four (five) charge 2/3 ($-1/3$) quarks and a vector-like third generation.
Possibilities beyond these models are also briefly considered.
\end{abstract}
\narrowtext
\section{Introduction}
\label{sec:int}

Nowadays it seems that in some sense the third generation may be different
from the other ones: although a heavy $top$ quark~\cite{top} can still
be  barely accommodated in the Standard Model it can bring some unexpected
features to the mass spectrum problem, also, possibly the
$bottom$ quark couples to the $Z^0$ with a
strength which is different from the strength of the $d$ and $s$
quarks~\cite{jr1}. The properties of
the $tau$ lepton and its neutrino can still bring up surprises\cite{tau}.
On the other hand, if the cross section $\sigma(p\bar p \to t\bar t+X)$
obtained by the CDF
Collaboration\cite{top} is in fact higher than the prediction of quantum
chromodynamics,
this may be a signature of new quarks.

In the model based on the gauge symmetry $SU(3)_c\otimes SU(3)_L\otimes
U(1)_N$ of Ref.~\cite{pp}, the effective $SU(2)\otimes U(1)$ model coincides
with the usual electroweak one. The three families belong to left-handed
doublets and right-handed singlets of $SU(2)$. Hence, all of them have,
at leading order, the
same couplings to the $W^\pm$ and $Z^0$ bosons. Also the extra quarks
in that model have exotic 5/3 and $-4/3$ charges. The lepton sector is
exactly the same of the Standard Model (SM).
Although this model coincides at low
energies with the usual electroweak model,
it explains
the fundamental questions: i) the family number and, ii) why
$\sin^2\theta_W<1/4$ is observed.
Therefore, it is possible from the last constraint, to compute an upper
limit to the mass scale of the $SU(3)$ breaking of about $3$ TeV~\cite{ng}.
This turns the
3-3-1 model an interesting possibility for the physics beyond the
Standard Model, in particular
if future experiences confirm in more detail an $SU(2)\otimes
U(1)$ model (for instance if the $Z\to b\bar b$ decay and
several $Z$-pole asymmetries confirm the
value expected in the model)
and no new quarks with charge 2/3 and $-1/3$ were found.

However, if in the future new quarks were observed
having the same charge than the quarks already
known or, if the third generation turns to be in
fact different from the other two generations
(say, with different interactions),
it will be necessary to consider modifications of the
original 3-3-1 model. For instance, a model with five charge $-1/3$
and four charge 2/3 quarks has already been considered in Ref.~\cite{mpp}.
There are, however, other representation contents: four charge $-1/3$ and
five charge $2/3$ quarks or, by the choice of the third generation in a
vector-like representation of the electroweak symmetry.
We will give below an example of
such a sort of model which is an extension of one of the models
proposed some years ago in Ref.~\cite{gp}.

Once we are convinced that theories based on
the 3-3-1 gauge symmetry are interesting possibilities for the
physics at the TeV range, we must study how the
basic ideas of this sort of models can be generalized.

In (almost) all these models, which we recall that are
indistinguishable from the Standard Model at low energies, in order
to cancel anomalies the number of families ($N_f$) must be divisible
by the number of color degrees of freedom (3). Hence the simplest
alternative is $3=N_f$. By denoting $N_q$ and $N_l$ the
number of quark and lepton families, respectively, we will see that
the relation $N_q=N_l=N_f=3$ is a particular feature
of 3-$m$-1 models. When $n\not=3$,  $N_q$ and $N_l$ are still
related to each other but it is not necessary that
$N_q=N_l$ in order to have anomaly cancellation.

In this work we also want
to generalize these sort of models in several ways. Firstly, by
expanding the color degrees of freedom $(n)$ and the electroweak
sector $(m)$, i.e., we will consider models based on the gauge
symmetry
\begin{equation}
SU(n)_c\otimes SU(m)_L\otimes U(1)_N.
\label{g}
\end{equation}
In most of these extensions the anomaly cancellation occurs among all
generations together, and not
generation per generation. However, if a 3-3-1 model has the third
generation  not anomalous, in its extension also it will be so.

We will use the criterion that the values for $m$ in Eq.~(\ref{g})
are determined by the leptonic sector. It means that if each generation
is treated separately, $SU(4)$ is the highest symmetry group
to be considered in the electroweak sector. Thus, there is no room for
$SU(5)_L\times U(1)_N$ if we restrict ourselves to the case of leptons
with charges $\pm1,0$.

In the color sector, for
simplicity, in addition to the usual case of $n=3$, we will
comment the cases $n=4,5$. These extensions have been considered in
the context of the $SU(2)_L\otimes U(1)_Y$ model~\cite{f,fh}.

Next, models with left-right symmetry and/or
with horizontal symmetry are also considered. We discuss too
a $SU(6)$ grand unified theory in which one of these models may be
embedded.

We must stress that all extensions of 3-3-1 models have
flavor changing neutral currents (FCNC). However, up to now in all models of
this kind which have been considered in detail, it was always possible to
have, in the sector of the model which coincides with the observed one,
natural conservation of flavor in the neutral currents.
Hence, FCNC effects are restricted to the
exotic sectors of the models. The only exception is model B below
since in this case there are right-handed currents coupled to the $Z^0$
which do not conserve flavor but they involve
arbitrary right-handed mixing matrix. Since all extensions of the 3-3-1 model
that we will consider here have an $SU(3)_L$ subgroup, we think that the
suppression of the FCNC in the observed part of the particle spectrum is
a general feature of this kind of models. This is far from being an obvious
fact but it was showed in
several 3-3-1 models and recently in the 3-4-1 case. For details see
Refs.~\cite{mpp,dumm,su4}.

This work is organized as follows. In Sec.\ref{sec:mab} we consider
two new possibilities of 3-3-1 models. In Sec.\ref{sec:nm1} we
consider models with $n=3$, $m=3,4$ (Sec.\ref{subsec:341}). There we
will also discuss the cases for $n=4,5$ (Sec.\ref{subsec:n45}). In
Sec.\ref{sec:lrhs} we give general features of the extensions with
left-right symmetry (Sec.\ref{subsec:lr}) and with horizontal
symmetries (Sec.\ref{subsec:ho}). We also consider
(Sec.\ref{subsec:su6}) possible embedding in $SU(6)$. The last
section is devoted to our conclusions.
\section{Two 3-3-1 models}
\label{sec:mab}

Here, we will treat two interesting possibilities of 3-3-1 models
with the electric charge operator defined as
$2Q=\lambda_3+\lambda_8/\sqrt3+2N$. Both models have the same
gauge boson sector, they differ slightly in the scalar sector but they
are quite different in the fermion sector.
One of the models (model A)
has five charge $2/3$ quarks and four charge $-1/3$ ones; the other model
(model B) has four charge 2/3 and five charge $-1/3$
quarks and
the third generation in a vector-like representation of $SU(3)$. Model B
is an extension with three quark generations of one of the models put forward
in Ref.~\cite{gp}. In model A anomalies cancel out only among all
generations with each generation being anomalous. In model B only
the third generation is not anomalous.

 \subsection{ Model A}
\label{subsec:ma}
All leptons generations transform as triplets of $SU(3)$
\begin{equation}
\Psi_{aL}=\left(
\begin{array}{c}
\nu_a \\ l^-_a \\ E^-_a
\end{array}
\right)_L\sim ({\bf3},-2/3),\quad a=e,\mu,\tau;
\label{l1}
\end{equation}
while quarks transform as follows
\begin{equation}
Q_{iL}=\left(
\begin{array}{c}
u'_i \\ u_i \\ d_i
\end{array}
\right)_L\sim ({\bf3}^*,1/3),\quad i=1,2;
\quad
Q_{3L}=\left(
\begin{array}{c}
u_3 \\ d_3 \\ d_4
\end{array}
\right)_L\sim ({\bf3},0),
\label{q1}
\end{equation}
and all charged right-handed fields in singlets. Neutrinos remain
massless as long as no right-handed components are introduced.
We have omitted the color index.

In the quark sector the
phenomenological states in Eqs.~(\ref{q1}) are linear
combinations
of the mass eigenstates $(u,c,t,t',t'')$ and $(d,s,b,b')$ for the
charge $2/3$ and charge $-1/3$ sectors, respectively. Three of the
left-handed charge $2/3$ ($-1/3$) quark fields are part of $SU(2)$
doublets $u_{1,2,3}$ ($d_{1,2,3}$). The other fields $u'_{1,2}$ and
$d_4$ are in singlets of $SU(2)$.

Let us introduce three triplets of Higgs bosons
\begin{equation}
\begin{array}{cc}
\eta=\left(\begin{array}{c}
\eta^+_2\\ \eta_1^+ \\ \eta^0
\end{array}\right)\sim({\bf3}^*,2/3),&
\sigma=\left(\begin{array}{c}
\sigma^0_2 \\ \sigma_1^0 \\ \sigma^-
\end{array}\right)\sim({\bf3}^*,-1/3),
\end{array}
\label{gp1}
\end{equation}
and a third one $\sigma'$ transforming like $\sigma$.

The most general quark Yukawa couplings are
\begin{eqnarray}
-{\cal L}_Y &=& \sum_{i\alpha}
A_{i\alpha}\bar{Q}_{iL}{\cal D}_{\alpha R}\eta
+\sum_{i\beta}[B_{i\beta}\bar{Q}_{iL}\sigma
+B'_{i\beta}\bar{Q}_{iL}{\sigma'}] {\cal U}_{\beta R}
\nonumber \\ & &\mbox{}
+\sum_\alpha [E_{3\alpha}\bar{Q}_{3L} \sigma^*+
E'_{3\alpha}\bar{Q}_{3L}{\sigma'}^*]{\cal D}_{\alpha R}+
\sum_\beta F_{3\beta}\bar{Q}_{3L} {\cal U}_{\beta R}\eta^*+H.c.
\label{gp2}
\end{eqnarray}
where $i=1,2$, $\alpha=1,2,3,4$, $\beta=1,2,3,4,5$ and we have chosen the
basis ${\cal D}_{\alpha R}=d_{1,2,3,4R}$,
${\cal U}_{\beta R}=u_{1,2,3R}, u'_{1,2R}$; with
$\eta^*,\sigma^*$ the respective antitriplets and we have omitted $SU(3)$
indices.

Let us assume the following vacuum expectation values (VEVs):
\begin{equation}
\langle\sigma_1^0\rangle\not=0,
\;\langle\sigma_2^0\rangle=0,\quad \langle{\sigma'}_1^0\rangle=0,
\;\langle{\sigma'}_2^0\rangle\not=0;
\label{vev1}
\end{equation}
and we also assume that the mass scale characteristic of the $SU(3)$
symmetry is rather high:
\begin{equation}
\langle{\sigma'}_2^0\rangle\gg
\langle\sigma_1^0\rangle, \langle\eta^0\rangle.
\label{vev2}
\end{equation}

Before going on, let us consider the neutral currents coupled to $Z^0$.
We must determine which fields have the same couplings than in the
$SU(2)\otimes U(1)$ effective theory i.e., when
the condition in Eq.~(\ref{vev2}) is satisfied.
The photon field is given by
\begin{mathletters}
\label{az}
\begin{equation}
A_\mu=s_W\,(W^3_\mu+\frac{1}{\sqrt3}W^8_\mu)+\frac{1}{\sqrt3}\,
(3-4s_W^2)^{\frac{1}{2}}B_\mu.
\label{foton}
\end{equation}
The massive neutral bosons are
\begin{equation}
Z_\mu=c_W \,W^3_\mu-\frac{1}{\sqrt3}\, \tan\theta_W\,[s_W\,W^8_\mu+
(3-4s_W^2)^{\frac{1}{2}}B_\mu],
\label{z0}
\end{equation}
which correspond to the usual $Z^0$ and the heavier one
\begin{equation}
Z'_\mu= \frac{1}{\sqrt3c_W}[-(3-4s_W^2)^{\frac{1}{2}}W^8_\mu
+s_W\,B_\mu]
\label{zp}
\end{equation}
\end{mathletters}
where $s_W\equiv\sin\theta_W$, $c_W\equiv\cos\theta_W$ and
$\theta_W$ is the usual weak mixing angle. From the electric charge
definition we get
\begin{equation}
\frac{g'^2}{g^2}= \frac{3 s^2_W}{3-4s^2_W},
\label{ggp}
\end{equation}
hence, $\sin^2\theta_W<3/4$~\cite{mpp}.

The neutral current interactions of fermions ($\psi_i$)
can be written as usual
\begin{eqnarray}
{\cal L}^{NC}&=&-\frac{g}{2\cos\theta_W}\left[\sum_i L_i\bar \psi_{iL}
\gamma^\mu
\psi_{iL}+R_i \psi_{iR}\gamma^\mu\psi_{iR}\right]Z_\mu \nonumber \\ &=&
-\frac{g}{2\cos\theta_W}\sum_i\bar\psi_i\gamma^\mu(g^i_V-\gamma^5 g^i_A)
\psi_i Z_\mu,
\label{nc}
\end{eqnarray}
where $g^i_V\equiv\frac{1}{2}(L_i+R_i)$ and $g^i_A\equiv\frac{1}{2}(L_i-R_i)$.
We obtain for the charge $-1/3$ sector
\begin{mathletters}
\label{lrvad}
\begin{equation}
L_{d_1}=L_{d_2}=L_{d_3}=-1+\frac{2}{3}\,s^2_W,\quad
L_{d_4}=\frac{2}{3}\,s^2_W,
\label{ld1}
\end{equation}
\begin{equation}
R_{d_1}=R_{d_2}=R_{d_3}=R_{d_4}=\frac{2}{3}\,s^2_W.
\label{rd1}
\end{equation}
and,
\begin{equation}
g^{d_1}_{V}=g^{d_2}_V=g^{d_3}_V=-\frac{1}{2}+\frac{2}{3}\,s^2_W,
\quad g^{d_4}_V
=\frac{2}{3}\,s^2_W;
\label{vd1}
\end{equation}
\begin{equation}
g^{d_1}_A=g^{d_2}_A=g^{d_3}_A=-\frac{1}{2},\quad g^{d_4}_A=0.
\label{ad1}
\end{equation}
\end{mathletters}

Similarly, for the charge $2/3$ sector
\begin{mathletters}
\label{lrvau}
\begin{equation}
L_{u_1}=L_{u_2}=L_{u_3}=1-\frac{4}{3}\,s^2_W,
\quad L_{u'_1}=L_{u'_2}=-\frac{4}{3}\,s^2_W,
\label{lu}
\end{equation}
\begin{equation}
R_{u_1}=R_{u_2}=R_{u_3}=R_{u'_1}=R_{u'_2}=-\frac{4}{3}\,s^2_W,
\label{ru1}
\end{equation}
and
\begin{equation}
g^{u_1}_V=g^{u_2}_V=g^{u_3}_V=\frac{1}{2}-\frac{4}{3}\,s^2_W,\quad
g^{u'_1}_V=g^{u'_2}_V=-\frac{4}{3}\,s^2_W,
\label{vu2}
\end{equation}
\begin{equation}
g^{u_1}_A=g^{u_2}_A=g^{u_3}_A=\frac{1}{2},\quad
g^{u'_1}_A=g^{u'_2}_A=0.
\label{vau}
\end{equation}
\end{mathletters}

{}From Eqs.~(\ref{lrvad}) we see that $d_1,d_2$ and $d_3$ have the same
couplings to the $Z^0$
as the $d,s,b$ quarks in the standard electroweak model,
but $d_4$ has a pure vector coupling. In the charge $2/3$
sector we observe from Eqs.~(\ref{lrvau}) that $u_1,u_2$ and $u_3$ have
the same couplings of
the usual $u,c,t$ quarks in the Standard Model, while $u'_1,u'_2$ have
pure vector couplings to the $Z^0$.
The mass eigenstates will be denoted by
$u,c,t,t',t''$ and $d,s,b,b'$. Hence, if we avoid a general mixing in the
mass matrix we will implement a GIM mechanism~\cite{gim} in the model.

Finally, for leptons we have
\begin{mathletters}
\label{lrl}
\begin{equation}
L_{\nu_a}=1,\quad  L_{l_a}=-1+2\,s^2_W,\quad L_{E_a}=2\,s^2_W,
\label{a3}
\end{equation}
\begin{equation}
R_{\nu_a}=0,\quad R_{l_a}=R_{E_a}=2\,s^2_W.
\label{b3}
\end{equation}
or,
\begin{equation}
g^{\nu_a}_V=\frac{1}{2},\quad g^{l_a}_V=-\frac{1}{2}+2\,s^2_W,\quad
g^{E_a}_V=2\,s^2_W,
\label{c3}
\end{equation}
\begin{equation}
g^{\nu_a}_A=\frac{1}{2},\quad g^{l_a}_V=-\frac{1}{2},\quad
g^{E_a}_A=0.
\label{d3}
\end{equation}
\end{mathletters}
We see that neutrinos and $e^-,\mu^-,\tau^-$ have the same couplings to
the $Z^0$ than in the $SU(2)\otimes U(1)$ model. The heavy leptons $E_a$ have
vector-like
couplings. Lepton couplings with the $Z'^0$ conserve flavor in each sector:
$\nu_a$, $l^-_a$ and $E^-_a$. This is not
a surprise since lepton generations are treated democratically.

Knowing the neutral current couplings, given in Eqs.~(\ref{lrvad}) and
(\ref{lrvau}), in order to avoid a general mixing in the mass matrices
we will introduce the following discrete symmetries:
\begin{mathletters}
\label{ds}
\begin{equation}
\label{ds1}
d_{1,2,3R}\to d_{1,2,3R};\;d_{4R}\to -d_{4R},\quad
u_{1,2,3R}\to u_{1,2,3R};\;  u'_{1,2R},\to  -u'_{1,2R},\;
\end{equation}
\begin{equation}
Q_{1,2,3L}\to Q_{1,2,3L}, \quad
\eta,\sigma\to \eta,\sigma,\quad \sigma'\to -\sigma'.
\label{ds2}
\end{equation}
\end{mathletters}

For leptons the discrete symmetries are
\begin{equation}
\Psi_{aL}\to\Psi_{aL},\quad l^-_{aR}\to l^-_{aR},\quad E^-_{aR}\to -E^-_{aR}.
\label{dsleptons}
\end{equation}

The quark mass terms have the form
\begin{equation}
\bar{\cal D}_{\alpha L}M^D_{\alpha\alpha'}{\cal D}_{\alpha' R},\quad
\bar{\cal U}_{\beta L}M^U_{\beta\beta'}{\cal U}_{\beta' R}.
\label{mm1}
\end{equation}
With the discrete symmetries in Eq.~(\ref{ds}) the mass matrices become
\begin{equation}
M^U=\left(
\begin{array}{cc}
M_1^U & 0 \\
0 & M_2^U
\end{array}
\right),
\label{mm2}
\end{equation}
for the charge $2/3$ sector, and
\begin{equation}
M^D=\left(
\begin{array}{cc}
M_1^D & 0 \\
0 & 1
\end{array}
\right),
\label{mm3}
\end{equation}
for the charge $-1/3$ one. $M^U_2$ and $M^D_1$ are
arbitrary $3\times 3$ matrices in the basis $u_1,u_2,u_3$ and $d_1,d_2,d_3$,
respectively; $M^U_1$ is an arbitrary $2\times 2$ matrix in the basis
$u'_1,u'_2$. The unit matrix in $M^D$ means that $d_4$ does not mix with the
other quarks of charge $-1/3$.
The mass matrices in Eqs.~(\ref{mm1})--(\ref{mm3}) can be
diagonalized. In terms of the mass eigenstates they become
\begin{equation}
\bar D_L \hat{M}^D D_R,\quad \bar U_L \hat{M}^U U_R
\label{me}
\end{equation}
where $\hat M^U=\mbox{diag}(m_u,m_c,m_t,m_{t'},m_{t''})$,
and $\hat M^D=\mbox{diag}(m_d,m_s,m_b)$ and $U$ and $D$ denote
$(u,c,t,t',t'')$ and $(d,s,b,b')$, respectively. Note that in fact,
the mixing occurs
among $u_1,u_2,u_3$; $u_4, u_5$ and  $d_1,d_2,d_3$ while $d_4$ does not mix
and it is in fact the $b'$ quark.

The mass matrices in Eqs.(\ref{mm2}) and (\ref{mm3}) are diagonalized by
performing the transformations
\begin{equation}
 {\cal U}_L=V^U_L U_L,\quad {\cal U}_R=V^U_R U_R, \quad
{\cal D}_L=V^D_L D_L,\quad {\cal D}_R=V^D_R D_R,
\label{utdu}
\end{equation}
with

\begin{mathletters}
\label{mv}
\begin{equation}
\begin{array}{cc}
V^U_L=\left(
\begin{array}{cc}
V^U_{1L} & 0 \\
0 & V^U_{2L}
\end{array}
\right),   &
V^U_{1R}=\left(\begin{array}{cc}
V^U_{1R} & 0 \\
0 & V^R_{2R}
\end{array}
\right),
\end{array}
\label{mvu}
\end{equation}
\begin{equation}
\begin{array}{cc}
V^D_L=\left(
\begin{array}{cc}
V^D_{1L} & 0 \\
0 & 1
\end{array}
\right),   &
V^U_R=\left(\begin{array}{cc}
V^D_{1R} & 0 \\
0 & 1
\end{array}
\right).
\end{array}
\label{mvd}
\end{equation}
\end{mathletters}
$V^D_{1L,R}$, $V^U_{1L,R}$ are unitary $3\times 3$ matrices and
$V^U_{2L,R}$ $2\times 2$ ones.

Next, let us consider the interactions in the quark sector.
We have the currents
\begin{eqnarray}
{\cal L}_q&=&-\frac{g}{\sqrt2}\left(\bar u'_{iL}\gamma^\mu u_{iL}X^0_\mu-
\bar u'_{iL}\gamma^\mu d_{iL} V^+_\mu+
\bar u_{iL}\gamma^\mu d_{iL}W^+_\mu \right.\nonumber \\ & & \mbox{}
\left.+\bar u_{3L}\gamma^\mu d_{3L} W^+_\mu + \bar u_{3L}\gamma^\mu d_{4L}
V^+_\mu + \bar d_{3L}\gamma^\mu d_{4L} X^0_\mu\right) + H.c.
\label{ccq}
\end{eqnarray}

In particular, the interaction with the $W^+$ boson can be written as usual
with the mixing matrix defined as $V_{KM}=V_L^{U\dagger}V_L^D$. On the other
hand we have the currents coupled to the $V^+$
\begin{equation}
{\cal L}^{CC}_{qV}=-\frac{g}{\sqrt2}(\bar u_{1}\, \bar u_{2}\,
\bar u_{3}\, \bar u'_{1}\,\bar u'_{2})_L \Delta\gamma^\mu
\left(
\begin{array}{c}
d_{1} \\ d_{2} \\ d_{3} \\ d_{4} \\ 0
\end{array}
\right)_LV^+_\mu+H.c.,
\label{cce1}
\end{equation}
where $\Delta$ is a $5\times 5$ matrix with $-\,\Delta_{41}=-\,\Delta_{52}=
\Delta_{34}=1$ and all other elements vanish. In terms of the
mass eigenstates we can write Eq.~(\ref{cce1}) as
\begin{equation}
{\cal L}^{CC}_{qV}=-\frac{g}{\sqrt2}(\bar u\;\; \bar c \;\;\bar t\;\;
\bar t' \;\;\bar t'')_L K \gamma^\mu
\left(
\begin{array}{c}
d \\ s \\ b \\ b'  \\ 0
\end{array}
\right)_LV^+_\mu+H.c.,
\label{cce2}
\end{equation}
where $K$ is defined as
\begin{equation}
K\equiv {V^U_L}^\dagger\Delta V^D_L,
\label{k}
\end{equation}
being $V^U_L, V^D_L$ the matrices in Eqs.~(\ref{mvu}) and (\ref{mvd}).
In these sort of models it is not interesting to define $V^U_{L}$ as
being the unit matrix as it is usually done in the $SU(2)\otimes U(1)$ model.
This is
because this matrix appears also in the neutral currents with the extra $Z'^0$
present in the model~\cite{dumm}. So we do not assume that the charge 2/3
mass eigenstates
appear unmixed.  There are similar interactions with the $X^0$ boson but
in this case there are mixture involving the matrices $V^U_{1L}$ and
$V^U_{2L}$ in the charge 2/3 sector and $V^D_{1L}$ in the charge $-1/3$ one.
We will return to this point later.

In the lepton sector, neutrinos and the usual charged leptons have the
same couplings to the $W^+$ boson
as in the $SU(2)\otimes U(1)$ model,
\begin{equation}
{\cal L}_l=-\frac{g}{2}\sum_a\left[\bar\nu_{aL}\gamma^\mu  l^-_{aL}W^+_\mu+
\bar\nu_{aL}\gamma^\mu E^-_{aL}V^+_\mu +\bar l_{aL}\gamma^\mu E^-_{aL}X_\mu
\right]
+H.c.,
\label{ccla}
\end{equation}

The charged
leptons get a mass via the interaction with the $\sigma^*$ and $\sigma'^*$
scalars. With discrete symmetries in Eq.~(\ref{dsleptons}) the Yukawa
interactions are
\begin{equation}
-{\cal L}_{lY}=\sum_{ab}\bar\Psi_{aL}\left[h_{ab}l_{bR}\sigma^*+h'_{ab}E_{bR}
{\sigma'}^*\right]+H.c.,
\label{yula}
\end{equation}
$h_{ab},h'_{ab}$ are arbitrary $3\times 3$ matrices and neutrinos remain
massless if right-handed neutrinos are not introduced.
We can define the neutrino fields in such a way that there is not mixing in
the $W^+$--interactions but there are mixings in the $V^+,X^0$ ones.

In the lepton sector we have no mixing among $l^-_a$ and $E^-_a$ in the
mass matrix since the discrete symmetries in (\ref{dsleptons}) forbid it.
So there is not flavor violation in Higgs-boson couplings.

\subsection{Model B}
\label{subsec:mb}
This is an extended version of the model proposed some years ago
by Georgi and Pais~\cite{gp}. Here
we have to consider a third quark generation since today there is evidence
of the existence of a $t$ quark~\cite{top}. As we will see later,
the phenomenology of this model is rather different from that of
Georgi and Pais's model in both sectors, quarks and leptons.

Let us first consider leptons. This is the same of Ref.~
\cite{gp} with four antitriplets $({\bf3}^*,-1/3)$
\begin{mathletters}
\label{l2}
\begin{equation}
\begin{array}{cccc}
\Psi_{eL}:\;\left(\begin{array}{c}
\nu^c_e\\ \nu_e \\ e^-
\end{array}\right)_L,& \Psi_{\mu L}:\;\left(\begin{array}{c}
\nu^c_\mu  \\ \nu_\mu \\ \mu^-
\end{array}\right)_L;  &
\Psi_{\tau L}:\;\left(\begin{array}{c}
\nu^c_\tau \\ \nu_\tau \\ \tau^-
\end{array}\right)_L,  &
\Psi_{TL}:\;\left(\begin{array}{c}
\nu^c_T\\ \nu_T \\ T^-
\end{array}\right)_L,
\end{array}
\label{lmb1}
\end{equation}
and two other antitriplets with $({\bf3}^*,2/3)$
\begin{equation}
\begin{array}{cc}
\Psi'_{eL}:\quad\left(\begin{array}{c}
e^+\\ \tau^+ \\ N^0_1
\end{array}\right)_L,&
\Psi'_{\mu L}:\quad\left(\begin{array}{c}
\mu^+ \\ T^+ \\ N^0_2
\end{array}\right)_L,
\end{array}
\label{lmb2}
\end{equation}
\end{mathletters}
and two neutral singlets $(N^0_{1L})^c,(N^0_{2L})^c$.

The quark fields of the first two generations
(suppressing the color indices) are in two left-handed triplets $({\bf3},0)$
\begin{equation}
Q_{iL}=\left(\begin{array}{c}
u_i \\ d_i \\ d'_i
\end{array}\right)_L\!,\;i=1,2
\label{q2}
\end{equation}
and the right-handed components in singlets $u_{iR}\sim({\bf1},2/3)$
and $d_{iR},d'_{iR}\sim({\bf1},-1/3)$. Finally, the third quark
generation transforms in a vector-like representation
\begin{equation}
\begin{array}{cc}
Q_{3L}=\left(\begin{array}{c}
u_4 \\ u_3\\ d_3
\end{array}\right)_L\sim({\bf3}^*,1/3), &
Q_{3R}=\left(\begin{array}{c}
u_4 \\ u_3\\ d_3
\end{array}\right)_R\sim({\bf3}^*,1/3).
\end{array}
\label{q3}
\end{equation}
The scalar fields are those in Eq.~(\ref{gp1}) but now we can introduce
a singlet neutral scalar $\phi$ with VEV $\langle \phi\rangle\not=0$.

As we said before, the gauge bosons are the same of model A.
In particular, the neutral ones are given in Eqs.~(\ref{az}).
Thus, in this model we get the neutral current couplings defined in
Eq.~(\ref{nc}),

\begin{mathletters}
\label{x1}
\begin{equation}
L_{d_1}=L_{d_2}=L_{d_3}=-1+\frac{2}{3}\,s^2_W, \quad
L_{d'_1}=L_{d'_2}=\frac{2}{3}\,s^2_W,
\label{c1}
\end{equation}
\begin{equation}
R_{d_1}=R_{d_2}=\frac{2}{3}
\sin^2\theta_W,\quad R_{d_3}=
-1+\frac{2}{3}\,s^2_W,\quad
R_{d'_1}=R_{d'_2}=\frac{2}{3}\,s^2_W,
\label{d1}
\end{equation}
or,
\begin{equation}
g^{d_1}_V=g^{d_2}_V=g^{d_3}_V=-\frac{1}{2}+\frac{2}{3}\,s^2_W,
\quad  g^{d'_1}_V=g^{d'_2}_V=\frac{2}{3}\,s^2_W,
 \label{c2}
 \end{equation}
 \begin{equation}
g^{d_1}_A=g^{d_2}_A=-\frac{1}{2},\quad g^{d_3}_A= g^{d'_1}_A=g^{d'_2}_A=0,
\label{d2}
\end{equation}
\end{mathletters}
for the charge $-1/3$ quarks. For the charge $2/3$ sector we have
\begin{mathletters}
\label{oba}
\begin{equation}
L_{u_1}=L_{u_2}=L_{u_3}=1-\frac{4}{3}\,s^2_W, \quad
L_{u_4}=-\frac{4}{3}\,s^2_W,
\label{a1}
\end{equation}
\begin{equation}
R_{u_1}=R_{u_2}=-\frac{4}{3}\,s^2_W,
\quad R_{u_3}=1-\frac{4}{3}\,s^2_W,\quad
R_{u_4}=-\frac{4}{3}\,s^2_W;
\label{b1}
\end{equation}
\end{mathletters}
or
\begin{mathletters}
\label{oba2}
\begin{equation}
g^{u_1}_V=g^{u_2}_V=g^{u_3}_V=\frac{1}{2}-\frac{4}{3}\,s^2_W, \quad
g^{u_4}_V=-\frac{4}{3}\,s^2_W,
\label{a2}
\end{equation}
\begin{equation}
g^{u_1}_A=g^{u_2}_A=\frac{1}{2}, \quad g^{u_3}_A=g^{u_4}_A=0.
\label{b2}
\end{equation}
\end{mathletters}

In this model, for leptons we have
\begin{mathletters}
\label{lrl2}
\begin{equation}
L_{\nu_a}=1,\quad L_{l_a}=-1+2\,s^2_W,\quad L_{N_1}=L_{N_2}=-1,
\label{a4}
\end{equation}
\begin{equation}
R_{\nu_a}=0,\;\;R_{e}=R_{\mu}=2\,s^2_W,\quad R_{\tau}=R_{T}
=-1+2\,s^2_W, \quad R_{N_1}=R_{N_2}=0.
\label{b4}
\end{equation}
where $\nu_a=\nu_e,\nu_\mu,\nu_\tau,\nu_T$ and $l_a=e,\mu,\tau,T$.
We see that neutrinos, electron and muon have the same couplings than in
the Standard Model, $N_i$ have right-handed couplings,
and the lepton $\tau$ and $T$ have both only vector couplings:
\begin{equation}
 g^{\nu_a}_V=\frac{1}{2}, \quad g^{e}_V=g^{\mu}_V-\frac{1}{2}+2\,s^2_W,\quad
 g^\tau_V=g^T_V=-1+2\,s^2_W, \quad g^{N_1}_V=g^{N_2}_V=-\frac{1}{2},
\label{vlb}
\end{equation}
\begin{equation}
g^{\nu_a}_A=\frac{1}{2},\quad g^{e}_A=g^{\mu}_A=0,\quad g^\tau_A=
g^T_A=0,\quad g^{N_1}_A= g^{N_2}_A=-\frac{1}{2}.
\label{alb}
\end{equation}
\end{mathletters}

Next, we will introduce the following discrete symmetries
\begin{mathletters}
\label{dsp}
\begin{equation}
\label{dsp1}
d_{1,2,3R}\to d_{1,2,3R},\; d'_{1,2R}\to -d'_{1,2R} ,\quad
u_{1,2,3R}\to u_{1,2,3R},\; u_{4R}\to -u_{4R};
\end{equation}
\begin{equation}
Q_{1,2L},\, Q_3\to Q_{1,2L}, \, Q_3; \quad
\eta,\sigma\phi\to \eta,\sigma,\phi;\quad \sigma'\to -\sigma'.
\label{dsp2}
\end{equation}
\begin{equation}
\Psi_{eL}, \Psi_{\mu L}\to \Psi_{eL}, \Psi_{\mu L},\quad
\Psi_{\tau L},\Psi_{TL}, \Psi'_{eL}, \Psi'_{\mu L}\to
-\Psi_\tau,-\Psi_T, -\Psi'_{eL}, -\Psi'_{\mu L},
\label{denovo}
\end{equation}
\begin{equation}
N_{1R}, N_{2R}\to -N_{1R},-N_{2R}.
\label{dn2}
\end{equation}
\end{mathletters}

As we said before, the Higgs multiplets are also the same that in the
previous model given in Eq.~(\ref{gp1}) but we can also introduce the
singlet $\phi$.
Hence, the most general Yukawa couplings
compatible with the symmetries in Eq.~(\ref{dsp}) are
\begin{eqnarray}
-{\cal L}_{qY} &=& \sum_{ij} \bar{{Q}_{iL}}\,[ A_{ij} u_{jR}\eta^*
+B_{ij}d_{j R}\sigma^*]
+\sum_{i\alpha}B'_{i\alpha} \bar{{Q}_{iL}}\,d'_{\alpha R}
{\sigma'}^*+\lambda\,\bar{Q_{3L}}\,Q_{3R}\,\phi
\nonumber \\ & &\mbox{} +\sum_j\bar{Q_{3L}}\,[E_{3j}u_{jR}\sigma+
D_{3j}d_{jR}\eta]+E'\bar{Q_{3L}}\,u_{4R}\sigma'+H.c.
\label{gp4}
\end{eqnarray}
Where $i,j=1,2,3$, $\alpha =1,2$ and $\eta^*,\sigma^*$ are the respective
antitriplets. (The $\bar\lambda$ matrices used in this work
are defined in the Appendix of Ref.~\cite{mpp}.)

{}From Eq.~(\ref{gp4}) we obtain mass matrices like in Eqs.~
(\ref{mm2}) and (\ref{mm3}) but now a $4\times 4$ matrix for the charge
2/3 sector and a $5\times 5$ one for the charge $-1/3$ sector. Instead
of the unitary matrices in Eqs.~(\ref{mvd}) and (\ref{mvu}) we get

\begin{mathletters}
\label{mvmb}
\begin{equation}
\begin{array}{cc}
V^D_L=\left(
\begin{array}{cc}
V^D_{1L} & 0 \\
0 & V^D_{2L}
\end{array}
\right),   &
V^D_R=\left(\begin{array}{cc}
V^D_{1R} & 0 \\
0 & V^R_{2R}
\end{array}
\right)
\end{array}
\label{mvdmb}
\end{equation}
\begin{equation}
\begin{array}{cc}
V^U_L=\left(
\begin{array}{cc}
V^U_{1L} & 0 \\
0 & 1
\end{array}
\right),   &
V^U_R=\left(\begin{array}{cc}
V^U_{1R} & 0 \\
0 & 1
\end{array}
\right).
\end{array}
\label{mvumb}
\end{equation}
\end{mathletters}
where $V^D_{1L,R},V^U_{1L,R}$ are unitary $3\times 3$ matrices and
$V^D_{2L,R}$ are $2\times 2$ ones.

Hence, we get a $2\times 2$ mass
matrix for $d'_\alpha$ quarks
depending only on $\langle{\sigma'}_2^0\rangle$; $3\times 3$ mass matrices for
$d_i$'s and $u_i$'s. Both matrices have contributions from the three Higgs
fields $\eta,\sigma$ and  $\phi$.
The fourth charge $2/3$ quark, $u_4$, get a mass
$m_{t'}=\lambda\langle\phi\rangle+E'\langle{\sigma'}_2^0\rangle$.

{}From the mass
matrices coming from Eq.~(\ref{gp4}) we obtain mixing among each of the
three sectors $u_i,d_i$ and $d'_i$, but $u_4$ does not mix at all.
Thus, $u_i$ have
the respective mass eigenstates $u,c,t$.
In the charge $-1/3$ sector
the mixing occurs among $d_i$ with the mass eigenstates denoted as usual
$d,s,b$ and among
the $d'_i$ sector with the respective mass eigenstates denoted by $s',b'$.

Finally, let us consider the lepton-scalar couplings.
The discrete symmetry in Eqs.~(\ref{denovo}) and (\ref{dn2}) allows
the following Yukawa interactions
\begin{equation}
-{\cal L}_{lY}=\sum_{ab}\epsilon\, \Gamma_{ab}
\overline{(\Psi_{aL})^c} \,\Psi_{bL}\sigma+\sum_{db}\epsilon\,\Gamma'_{db}
\overline{(\Psi_{dL})^c} \,\Psi'_{b L}\sigma'+
\sum_{bi}H_b\overline{\Psi'_{bL}}\,N^0_{iR}\eta+H.c.
\label{ul}
\end{equation}
where $a,b=e,\mu$; $d=\tau,T$; $i=1,2$; $\Psi^c$ is the charge
conjugated field and $\epsilon$ is the completely antisymmetric $SU(3)$
tensor. Recall that the scalar fields must obey the discrete symmetry
in Eq.~(\ref{dsp2}).

Charged and non-hermitian neutral currents are
\begin{eqnarray}
{\cal L}_q&=&-\frac{g}{\sqrt2}\left[\sum_i\left(\bar{u}_{iL}\gamma^\mu
d_{iL}W^+_\mu +\bar{u}_{iL}\gamma^\mu d'_{iL}V^+_\mu
+\bar d_{iL}\gamma^\mu d'_{iL}X^0_\mu\right)
+\bar u_{4L}\gamma^\mu u_{3L} X^0_\mu
-\bar{u}_{4L}\gamma^\mu d_{3L} V^+_\mu\right. \nonumber \\ & &\mbox{}
\left.+\bar{u}_{3L}\gamma^\mu d_{3L}W^+_\mu+
 \bar u_{4R}\gamma^\mu u_{3R}X^0_\mu-
\bar u_{4R}\gamma^\mu d_{3R}V^+_\mu+\bar u_{3R}\gamma^\mu d_{3R}W^+_\mu
\right]+H.c
\label{ccb}
\end{eqnarray}
for quarks, and
\begin{eqnarray}
{\cal L}_l&=& \frac{g}{\sqrt2}\left[ \sum_a\left(\bar\nu^c_{aL}\gamma^\mu
\nu_{aL}X^0_\mu-
\bar\nu^c_{aL}\gamma^\mu l^-_{aL}V^+_\mu+\bar\nu_{aL}\gamma^\mu l^-_{aL}
W^+_\mu\right)  +\bar e^+_L\gamma^\mu\tau^+_L
X^0_\mu-\bar e^+_L\gamma^\mu N^0_{1L}V^+_\mu
\right. \nonumber \\ & &\mbox{}\left.
+\,\bar \tau^+_L\gamma^\mu N^0_{1L}W^+_\mu
+\bar \mu^+_L\gamma^\mu T^+_LX^0_\mu-\bar\mu^+_L
\gamma^\mu N^0_{2L}V^+_\mu+\bar T^+_L\gamma^\mu N^0_{2L}W^+_\mu\right]+H.c.
\label{cclb}
\end{eqnarray}
for leptons. As long as neutrinos remain massless there is not mixing
in the charged currents coupled to $W^+$. However, there is mixing
in the currents coupled to $V^+$ and $X^0$. Notice that the right-handed
currents coupled to $W^+$ involve the charged leptons $\tau$ and $T$ and
the heavy neutral fermions $N_{1,2}$.

After neutrinos getting mass
through radiative corrections, mixing will appear in the interactions
with $W^+$. Recall that the scalar $\eta^+_2$ is an $SU(2)$ singlet and there
are three $SU(2)$ doublets in the scalar multiplets given in Eq.~(\ref{gp1}),
hence the Zee mechanism for generating neutrino masses may be
implemented in this models~\cite{zee}.

The charged currents coupled to the $W^+$ boson can be written, in
the quark sector, as
\begin{equation}
{\cal L}^{CC}_{qW}=-\frac{g}{\sqrt2}(\bar u_1 \; \bar u_2\; \bar u_3)_L
\gamma^\mu\left( \begin{array}{c}
d_1 \\ d_2 \\ d_3 \end{array}
\right)_LW^+_\mu -\frac{g}{\sqrt2}\bar u_{3R}\gamma^\mu d_{3R}W^+_\mu+H.c.
\label{wb}
\end{equation}
The left-handed currents in Eq.~(\ref{wb}) can be parametrized
in terms of mass eigenstates and Kobayashi-Maskawa mixing matrix,
$V_{KM}=V^{U\dagger}_LV^D_L$.
The right-handed current in Eq.~(\ref{wb}) can be written in terms of
the mass eigenstates as follows
\begin{equation}
{\cal L}^{CC}_{qWR}=-\frac{g}{\sqrt2}(\bar u\; \; \bar c\;\; \bar t)_R
V^{U\dagger}_R
\gamma^\mu\tilde\Delta \,V^D_R \left( \begin{array}{c}
d \\ s \\ b \end{array} \right)_R W^+_\mu+H.c.
\label{wbr}
\end{equation}
where $\tilde\Delta=\mbox{diag}(0,0,1)$. The other matrices $V^U_L,V^D_L$
also survive in the interactions involving the bosons $V^+$ and $X^0$.

Notice that, besides the neutral currents coupled to the $Z^0$
given in Eq.(\ref{nc}) we have additional right-handed couplings
$[\bar u_{3R}\gamma^\mu u_{3R}-\bar d_{3R}\gamma^\mu d_{3R}]Z^0_\mu$,
or, written in terms of the mass eigenstates
\begin{equation}
{{\cal L}''}^{NC}={\cal L}^{NC}+ {{\cal L}'}^{NC},
\label{ncp}
\end{equation}
with ${\cal L}^{NC}$ being parametrized like in Eq.~(\ref{nc}) and
the coefficients $L,R$s or $V,A$s being defined as in the Standard Model,
and  with
\begin{equation}
{{\cal L}'}^{NC}_U=-\frac{g}{2\cos\theta_W}(\bar u\;\; \bar c\;\; \bar t)_R
\gamma^\mu
{U^U_R}^\dagger\tilde\Delta \,U^U_R\left(
\begin{array}{c}
u \\ c \\ t
\end{array}
\right)_R\,Z^0_\mu,
\label{ru2}
\end{equation}
for the charge 2/3 sector, and
\begin{equation}
{{\cal L}'}^{NC}_D=+\frac{g}{2\cos\theta_W}(\bar d\;\; \bar s\;\; \bar b)_R
\gamma^\mu {U^D_R}^\dagger\tilde \Delta\, U^D_R \left(
\begin{array}{c}
d \\ s \\ b
\end{array}
\right)_R\,Z^0_\mu.
\label{rd2}
\end{equation}
for the charge $-1/3$ one.
There are similar currents coupled to the $V^+,X^0,Z'^0$ bosons. Although
Eqs.~(\ref{ru2}) and (\ref{rd2}) are FCNC, all flavors
have the same dependence on the weak mixing
angle. This is not the case of the neutral currents coupled to $Z'^0$, here
each flavor has a different dependence on that angle. In the leptonic sector
we have the GIM mechanism at tree level in the neutral currents coupled
to the $Z^0$ and $Z'$ (as long as the symmetries in Eqs.~(\ref{denovo})
and (\ref{dn2})
are preserved) but there are FCNC in the couplings to the $X^0$.
Mixing between $e\leftrightarrow\mu$ and $\tau\leftrightarrow T$
appear in the current coupled to $X^0$ and those coupled to
$V^+$ induce transitions $l_a\leftrightarrow N_i$ which
are sensible on the Cabibbo-like mixing in the charged leptons.

Unlike the model of Ref.~\cite{gp}, in the present model all Higgs
bosons couple to quarks even the scalar with a large VEV. Another
difference between our model B and that of Ref.~\cite{gp} is that in the
latter one, there is a mixing between $d,s$ and between $b,b'$. In our case
the mixing is as usual among $d,s,b$, but $b'$ has no mixing with
other charge $-1/3$ quarks, at least as long as the discrete symmetries
in Eq.~(\ref{dsp}) are preserved. In fact, the mass matrices in model B
are different from those of the model of Ref.~\cite{gp}.
 We can see these discrete symmetries
only as an indication of which ones are the dominant mixings. Eventually,
we could allow them to be broken.

Notice also that in both models, A and B, the extra quarks are very heavy
since they get mass through the larger VEV $\langle{\sigma'}^0\rangle$.

We can also build a model in which there are two quark generations
transforming as $({\bf3}^*,1/3)_L$ and one as $({\bf3},0)_{L+R}$. In
this case there are two leptonic antitriplets $({\bf3},1/3)$
and four ones transforming as $({\bf3},-2/3)$. In this case it is
necessary, however, to include right-handed charged leptons in singlets.

\section{Models with extended color and electroweak sectors}
\label{sec:nm1}
\subsection{Models with extended electroweak sector}
\label{subsec:341}
First, let us consider $n=3$ models. When $m=2,3$ we have
the Standard Model and the 3-3-1 models respectively.
Next, there is a 3-4-1 model in which the electric
charge operator is defined as
\begin{equation}
Q=\frac{1}{2}\left(\lambda_3-\frac{1}{\sqrt3}\lambda_8-\frac{2}{3}{\sqrt6}
\lambda_{15}\right)+N,
\label{q}
\end{equation}
where the $\lambda$-matrices are~\cite{da},
\[\lambda_3=\mbox{diag}(1,-1,0,0),\;\lambda_8=\left(\frac{1}{\sqrt{3}}\right)
\mbox{diag}(1,1,-2,0),
\;\lambda_{15}=\left(\frac{1}{\sqrt{6}}\right)\mbox{diag}(1,1,1,-3).\]

Leptons transform as $({\bf1},{\bf4},0)$,
two of the three quark families, say $Q_{iL},\,i=1,2$, transform as
$({\bf3},{\bf4}^*,-1/3)$,
and one family, $Q_{3L}$, transforms as $({\bf3},{\bf4},+2/3)$
\begin{equation}
\psi_{aL}=\left(\begin{array}{c}
\nu_a\\
l_a\\
\nu^c_a\\
l^c_a
\end{array}\right)_L, \qquad
Q_{iL} = \left(
\begin{array}{c}
j_i\\
d'_i\\
u_i\\
d_i
\end{array}\right)_L,\qquad
Q_{3L} = \left(
\begin{array}{c}
u_3\\
d_3\\
u'_3\\
J
\end{array}\right)_L,
\label{lq}
\end{equation}
where $u'_3$ and $J$ are new quarks with charge $+2/3$
and $+5/3$ respectively;
$j_i$ and $d'_i$, $i=1,2$ are new quarks with charge $-4/3$ and
$-1/3$ respectively. We remind that in Eq.~(\ref{lq})
all fields are still symmetry eigenstates.
Right-handed quarks transform as singlets under $SU(4)$.

A model with $SU(4)_L$ symmetry and leptons transforming as in
Eq.~(\ref{lq}) was proposed by Voloshin some years
ago~\cite{voloshin}. In this context it can be
possible to understand the existence of neutrinos with large magnetic
moment and small mass. However in Ref.~\cite{voloshin} it was not
considered the quark sector.

Quark masses are generated by introducing the following Higgs
$SU(4)_L\otimes U(1)_N$ multiplets:
$\chi\sim({\bf4},-1),\rho\sim({\bf4},+1)$, $ \eta$ and
$\eta'\sim({\bf4},0)$.

In order to obtain massive charged leptons it is necessary to
introduce a $({\bf10}^*,0)$ Higgs multiplet, because the lepton mass
term transforms as
$\bar \psi^c_L\psi_L\sim ({\bf6}_A\oplus {\bf10}_S)$.
The ${\bf6}_A$ will leave some leptons massless and some others mass
degenerate. Therefore we will choose $H={\bf10}_S$.
Neutrinos remain massless at least at tree level but the charged
leptons gain mass. The corresponding VEVs are
$\langle\eta\rangle=(v,0,0,0)$, $\langle\rho\rangle=(0,w,0,0)$,
$\langle\eta'\rangle=(0,0,v',0)$,
$\langle\chi\rangle=(0,0,0,u)$,
and $\langle H\rangle_{42}=v''$ for the decuplet. In this way we have
that the symmetry breaking of the $SU(4)_L\otimes U(1)_N$ group down
to $SU(3)_L\otimes
U(1)_{N'}$ is induced by the $\chi$ Higgs. The $SU(3)_L\otimes U(1)_{N'}$
symmetry is broken down into $U(1)_{em}$ by the $\rho,\eta$, $\eta'$
and $H$ Higgs. As in the models of Sec.\ref{sec:mab}, it is necessary
to introduce some discrete symmetries which
ensure that the Higgs fields give a
quark mass matrix in the charge $-1/3$ and  $2/3$ sectors of the
direct sum form in order to avoid general mixing among
quarks of the same charge. In this case the quark mass matrices can
be diagonalized with unitary matrices which are themselves direct sum
of unitary matrices.

In fact, we have the symmetry breaking pattern, including the $SU(3)$
of color,
\begin{equation}\begin{array}{c}
SU(3)_c \otimes SU(4)_L \otimes U(1)_N \\
\downarrow \langle \chi \rangle\\
SU(3)_c \otimes SU(3)_L \otimes U(1)_{N'}\\
\downarrow \langle \eta'\rangle\\
SU(3)_c\otimes SU(2)_L \otimes U(1)_{N''}\\
\downarrow \langle x \rangle\\
SU(3)_c \otimes U(1)_{em}
\end{array} \label{sbp} \end{equation}
where $\langle x\rangle$ means $\langle \rho\rangle$, $\langle
\eta\rangle$, $\langle H\rangle$~\cite{su4}.

The electroweak gauge bosons of this theory consist of a
${\bf15}$ $W^i_\mu$, $i=1,...,15$ associated with $SU(4)_L$ and
a singlet $B_\mu$ associated with $U(1)_N$.

There are four neutral bosons: a massless $\gamma$ and three massive
ones: $Z,Z',Z''$. The lightest one, say the $Z$, corresponds to the
Weinberg-Salam-Glashow neutral boson. Assuming the approximation
$u\gg v'\gg v,v'',w$ the extra neutral bosons,
say $Z',Z''$, have masses which depend mainly on $u,v'$.

Concerning the charged vector bosons, as in the model of
Ref.~\cite{pp} there are doubly charged vector bosons and there are
doublets of $SU(2)$
$(X^+_\mu,X^0)$ and $(\bar X^0_\mu,X^-_\mu)$
which produce interactions like $\bar
\nu^c_{aL}\gamma^\mu l_{aL}X^+_\mu$ and
$\bar\nu^c_{aL}\gamma^\mu\nu_{aL} X^0_\mu$, as in
model I of Ref.~\cite{mpp}. We have also the
$V^\pm_{1,2}$ vector bosons with interactions like $\bar l^c_{aL}\gamma^\mu
\nu_{aL}V^+_{1\mu}$ and $\bar l^c_{aL}\gamma^\mu \nu^c_{aL}V^+_{2L}$. All
charged currents, including that ones coupled with quarks,
are given in Ref.~\cite{su4}.

\subsection{$n=4,5$ models}
\label{subsec:n45}
Let us consider now $n=4,5$ models. Although the $SU(3)_c$ gauge
symmetry is the best candidate for the theory of the strong
interactions, there is no fundamental reason
why the colored gauge group must be $SU(3)_c$. In fact, it is
possible to consider other Lie groups. In general we have the
possibilities $SU(n),\,n\geq 3$~\cite{so}.

In particular, models in which quarks transform
under the fundamental representations of $SU(4)_c$ and $SU(5)_c$ were
considered in Refs.~\cite{f} and \cite{fh}, respectively, in the context
of the SM. These models
preserve the experimental consistency of the SM at low
energies. For instance, in the $SU(5)_c\otimes SU(m)_L\otimes U(1)_N$
model a Higgs
field transforming as the ${\bf10}$ representation of $SU(5)_c$
breaks the symmetry as follows~\cite{fh}
\begin{equation}
\begin{array}{c}
SU(5)_c \otimes SU(m)_L \otimes U(1)_N\\
\downarrow \langle {\bf10} \rangle\\
SU(3)_c \otimes SU(2)'\otimes SU(m)_L\otimes U(1)_N.
\end{array} \label{sbp3} \end{equation}
Later the electroweak symmetry will be broken and the remaining
symmetry will be $SU(3)_c\otimes SU(2)'\otimes U(1)_{em}$ as in the
models with $m=3,4$ considered above. Notice that, due to the
relation between the color degrees of freedom and the number of
families, it is necessary to introduce four and five leptonic families
for $n=4$ and $n=5$ respectively if we assume that the
number of quark families
is still three. In general we have $N_l=\vert n(n_1-n_2)\vert$ where
$n_1$ and $n_2$ are the number of quark multiplets transforming as
${\bf m}$ and ${\bf m}^*$ respectively and $N_q=n_1+n_2$. If
$n_1>n_2$ leptons must transform as ${\bf m}^*$ and if $n_1<n_2$
leptons are assigned to ${\bf m}$. It is still possible to have
$N_q=N_l$. Assuming this condition (and $n_1>n_2$), for the
case of even $n$ i.e., $n=2p$, $p\geq 2$  we have
$n_1/n_2=(2p+1)/(2p-1)$; and for odd $n$ i.e., $n=2p+1,\;p\geq1$ we get
$n_1/n_2=(p+1)/p$. For $n=4$ the condition $N_q=N_l$
is satisfied if $n_1/n_2=5/3$. Analogously, for $n=5$
we have $n_1/n_2=3/2$.
It means that, if we let the number of quark families to be equal
to the number of the lepton families, the
minimal number of families is eight for $n=4$ and five for $n=5$.

On the other hand, if we maintain $N_q=3$ it is necessary,
as we said before, to introduce new lepton families.
Let us denote these additional
families by $(N_i,E_i,E^c_i)$ with $i=1$ for $n=4$; or $i=1,2$ when $n=5$.
The new charged leptons must be heavy enough in order to keep consistency
with phenomenology. Since the right-handed neutrinos, transforming as
singlets under the gauge group, do not contribute to the
anomaly, their number is not constrained by the requirement of
obtaining an anomaly free theory. Hence, we can introduce an arbitrary
number of such fields. When these singlets are added,
the $Z^0$ invisible width is always smaller than the prediction of the
minimal SM. In fact it has been shown that in this case
\cite{cecilia}
\begin{equation}
\Gamma(Z\to\mbox{neutrinos})\leq N_l\Gamma^0,
\label{width}
\end{equation}
where $N_l$ is the number of left-handed lepton families and $\Gamma^0$
is the standard width for one massless neutrino.
Hence, it will be always possible to choose the neutrinos's mixing
angles and masses in such
a way that the theoretical value in (\ref{width}) be consistent with
the experimental one~\cite{pdg,tau3}.

\section{Other possible extensions}
\label{sec:lrhs}
Other possibilities are models with left-right symmetry in the
electroweak sector
$SU(n)_c\otimes SU(m)_L\otimes SU(m)_R\otimes U(1)_N$
and also models with horizontal symmetries $G_H$ i.e.,
$SU(n)_c\otimes SU(m)_L\otimes U(1)_N\otimes G_H$.

\subsection{Left-right symmetries}
\label{subsec:lr}
In models with left-right symmetry the $V-A$ structure of weak
interactions is related to the mass difference between the left- and
right-handed gauge bosons, $W^\pm_L$ and $W^\pm_R$, respectively, as a
result of the spontaneous symmetry breaking~\cite{gs}.

This sort of models are easily implemented in the 3-3-1 context by
adding a new charged lepton $E$. For instance, in models with
left-handed leptons transforming as $(\nu_a,l^-_a,E^+_a)^T_L$ the
right-handed triplet is $(\nu_a,l^-_a,E^+_a)^T_R$. In the quark sector, the
left-handed components are as in Ref.~\cite{pp} and similarly the
right-handed components, in such a way
that anomalies cancel in each chiral sector. Explicitly, the charge
operator is defined as
\begin{equation}
Q=I_{3L}+I_{3R}+\frac{Y}{2}
\label{colr}
\end{equation}
where $I_{3L(3R)}$ and $Y/2$ are of the form $(1/2)\lambda_3$
and $-(\sqrt{3}/2)\lambda_8+N{\bf1}$, respectively, for the model of
Refs.~\cite{pp}.
The Higgs multiplet $({\bf3},{\bf3}^*,0)$ and its conjugate give
mass to all fermions but in order to complete the symmetry breaking
it is necessary to add more Higgs multiplets.

\subsection{Horizontal Symmetries}
\label{subsec:ho}
Particle mixture occurs in the Standard Model among particles which
are equivalent concerning their position in the gauge multiplets.
It was noted some years ago that it is possible to determine the weak
mixing angles in terms of the quarks masses, provided we assume that all
equivalent multiplets of the vertical gauge symmetry
transform in the same way under horizontal symmetries.
Therefore, the three families are put into a single
representation of the horizontal group~\cite{wz}.

That is, in the context of the SM the gauge symmetry in the
horizontal direction was considered as a transformation among the
left-handed doublets and among right-handed singlets.
At first sight, horizontal
symmetries are less interesting in the context of 3-3-1 models since
quark generations transform in a different way under $SU(3)_L\otimes U(1)_N$.

Apparently, the only possibility is the horizontal $G_H=SU(2)_H$ symmetry.
In this case there are no additional conditions for canceling gauge anomalies
since $SU(2)$ is a safe group. For instance, with $n=3$, $m=3,4$, the
three quark generations transform, in both left- and right- handed sectors,
in the following way: two of them as a doublet and the third one as a
singlet under $SU(2)_H$~\cite{fe,dss}. The same is valid for leptons but
in this case the three lepton triplets can transform as the adjoint
representation as well.

The horizontal gauge bosons and the extra Higgs bosons have to be heavier
than the $W$ bosons or very weakly coupled to the usual fermions in
order to suppress appropriately flavor changing neutral transitions in
both, quark and lepton sectors.

\subsection{Embedding in $SU(6)$}
\label{subsec:su6}
There are also the grand unified extensions of all the possibilities
we have treated above.
The group $SU(3)_c\otimes
SU(3)_L\otimes U(1)_N$ has rank 5 and it is a subgroup of
$SU(6)$. In the last group, it has been shown that the anomalies,
${\cal A}$, of ${\bf15}$ and ${\bf6}^*$ are such that ${\cal
A}({\bf15})=-2{\cal A}({\bf6}^*)$~\cite{bg}. Then, pairs of
${\bf15}$ and ${\bf15}^*$; ${\bf6}^*$ and ${\bf6}$~\cite{gg} and,
finally one ${\bf15}$ and two ${\bf6}^*$ are the smallest anomaly
free irreducible representations in $SU(6)$. On the other hand, the
representation ${\bf20}$ is safe.

Just as an example, let us consider the $SU(6)$ symmetry which is
a possible unified theory for model B.
Using the
notation of Ref.~\cite{slansky}, in the entry $({\bf a},{\bf b})_f(N)$,
${\bf a}$ is an irreducible representation of $SU(3)_c$ and ${\bf b}$
is an irreducible representation of $SU(3)_L$. The subindex $f$
means, in an obvious notation, the respective fields of the model and
the second parenthesis contains the value of the $U(1)_N$ generator
when acting on the states in the $({\bf a},{\bf b})$.

In model B there are $66$ degrees of freedom.
Left-handed leptons and three of the right-handed $d$-type
quarks are in ${\bf6}^*$:
\begin{equation}
{\bf6^*}_{j}=({\bf3}^*,{\bf1})_{d^c_{jL}}(+1/3)+({\bf1},
{\bf3}^*)_{j_L}(-1/3),
\label{6l}
\end{equation}
where $d^c_{jL}=d^c_{1L},d^c_{2L},d'^c_{1L},d'^c_{2L}$;
$\Psi_{jL}=\Psi_{eL},\Psi_{\mu L},\Psi_{\tau L}, \Psi_{TL}$
(see Eq.(\ref{lmb1})).
Two quark generations transforming as $({\bf3},{\bf3},0)$ and
the right-handed $u$-type quarks are in two ${\bf15}$
\begin{equation}
{\bf15}_{Q_{iL}}=({\bf3}^*,{\bf1})_{u^c_{iL}}(-2/3)
+({\bf1},{\bf3}^*)_{\Psi'_{iL}}(2/3)+({\bf3},{\bf3})_{Q_{iL}}(0),
\label{151}
\end{equation}
where $u^c_{iL}=u^c_{1L},u^c_{2L}$; $\Psi'_{iL}=\Psi'_{e L},
\Psi'_{\mu L}$ (see Eq.~(\ref{lmb2})).
The left-handed and right-handed quarks of the third generation are
in one ${\bf20}$
\begin{equation}
{\bf20}_{Q_{3L}}=({\bf1},{\bf1})_{N^c_{1L}}(0)+({\bf1},{\bf1})_{N^c_{2L}}(0)
+({\bf3},{\bf3}^*)_{Q_{3L}}(+1/3)+({\bf3}^*,{\bf3})_{Q^c_{3L}}(-1/3),
\label{152}
\end{equation}
with $N^c_{iL}$, $i=1,2$ the neutral leptons which are singlets of the 3-3-1
symmetry. Thus,
we have an anomaly free theory with the fields of the first two generations
in ${\bf6}^*$ and ${\bf15}$. The third generation is not anomalous.

Let us consider the prediction of the weak mixing angle, $\sin\theta_W$. In
$SU(N)$ theories we have
\begin{equation}
\sin^2\theta_W=\frac{\sum_a(t_{3a})^2}{\sum_a(Q_a)^2},
\label{theta}
\end{equation}
where $t_3$ is the third component of the weak isospin, $Q$ is the electric
charge and the sum extends over all fields in a given representation.
Hence, in $SU(6)$ we have the prediction that $\sin^2\theta_W=3/8$.
This is the same value of
the $SU(5)$ model~\cite{su5}. It is easy to verify that all representations
${\bf6}^*$, ${\bf15}$ and ${\bf20}$ in Eqs.~(\ref{6l})-(\ref{152}) give the
same answer as it must be. On the other hand, we recall that in
models A and B it holds that
$\sin^2\theta_W<3/4$~\cite{mpp}. Thus, the theory has a Landau pole when
$\sin^2\theta_W=3/4$. The theory might be, before getting
this pole, unified in an $SU(6)$ model.

However, it is not a trivial issue to show that the
unification in $SU(6)$ may
actually occur~\cite{fn2}. This is so, because in 3-3-1 the couplings
$\alpha_c$
and $\alpha_{3L}$ have $\beta_{c}>\beta_{3L}$.

Since the model have new particles, we may have to consider mass
threshold corrections for the
$\beta$-functions, since the new particles could have masses below
the unification energy scale, or even, we may not assume the decoupling
theorem. We recall that in the Standard Model with two
Higgs doublets the decoupling theorem~\cite{ac} must not be necessarily
valid, since there are physical effects proportional to
$m^2_{\mbox{Higgs}}$~\cite{dt}. Hence, it could be interesting to study the
way in which the masses of the extra Higgs and exotic
quarks in the model become large, as it has been done in the standard
model scenario for an extended Higgs sector~\cite{dt} or for
the mass difference between fermions of a multiplet~\cite{veltman}.
It means that there is no ``grand
desert'' if 3-3-1 models are realized in nature.

How can we study the embedding of the SM
in 3-3-1? The last model has fields which do not exist in the
minimal SM, but which are present in the same multiplet of 3-3-1 with the
known quarks. For instance, the quarks $J$'s have to be added to the
SM transforming as $({\bf3},{\bf1},Q_J)$ under the 3-2-1 factors. The
scalar and vector boson sectors of the SM have also to be extended
with new fields. Hence, we must add
scalar fields transforming as (i) four singlets $({\bf1},{\bf1},Y_S)$:
one with $Y_S=0$, one with $Y_S=1$ and two with $Y_S=2$,
(ii) four doublets $({\bf1},{\bf2},Y_D)$: one with $Y_D=-3$ and three
with $Y_D=1$;  finally, (iii)
one  triplet $({\bf1},{\bf3},-2)$. It is also necessary to add extra
vector bosons ($U^{++},V^+$) which transform as $({\bf1},{\bf2},3)$.
For this reason we believe that 3-3-1 models are not just an
embedding of the SM but an alternative to describe the same
interactions.
\section{Conclusions}
\label{sec:con}
The 3-3-1 symmetry is in fact an interesting extension of the standard
model. It gives answers to some questions put forward by the
later model and new physics could arise at
not too high energies, say in the TeV range.

In the previous sections we have examined two 3-3-1 models,
both of them
with extra heavy quarks and leptons, and also possible extensions.

What we want to do now is to
discuss briefly some possible phenomenological consequences concerning
models A and B dicussed in Secs. \ref{subsec:ma} and
\ref{subsec:mb}, respectively.

\noindent {\em i)} In the Higgs sector,
by using the gauge
invariance it is not possible to choose all VEVs to be real. Hence,
there is CP violation via scalars exchange. Since the quark mass matrices
receive contributions from two VEVs, there are also FCNC in
the Higgs bosons couplings but their effects could be suppressed by fine
tuning among some parameters \cite{mpp} or by heavy scalars.

\noindent {\em ii)} In model A, the left-handed quark mixing matrices
$V^U_L$ and $V^D_L$, defined in Eqs.~(\ref{utdu}) or (\ref{mv}),
survive in the Lagrangian. See for instance,
Eqs.~(\ref{cce2}) and (\ref{k}).
Mixings are also different in the interactions with $X^0_\mu$
and with $V^-_\mu$, as can be seen from Eq.~(\ref{ccq}).
This induces new sources of CP
violation since there are phases in these interactions which cannot be
absorbed. This also happens in model B. However, in this case even
the right-handed quark mixing matrices, $V^U_R$ and $V^D_R$, survive.
We recall that in the
Standard Model although the matrices $V^{U,D}_{L,R}$ are needed, after
the diagonalization of the quark mass matrices the only place in the
Lagrangian where
these matrices appear is in charged currents coupled to the $W^+$
and only in the form $V^{U\dagger}_LV^D_L$. In this case, $V^D_L$ is
identified with the usual Cabibbo-Kobayashi-Maskawa mixing matrix by
choosing $V^U_L={\bf1}$.
Since this matrix does
not appear in other places of the Lagrangian, this choice is enough.
This is not the case for all 3-3-1 models~\cite{dumm}.

\noindent {\em iii)} It is very well known that almost all $Z^0$-pole
observables
are in agreement with the Standard Model predictions\cite{pdg}.
There are, however,
two of these observables which seem not to agree with the model's predictions:

a) the first one concerns the heavy quark production rates
$R_f=\Gamma(Z^0\to f\bar f)/\Gamma(Z\to hadrons)$,
which have been measured for $c$ and $b$ quarks. Considering $R_c$
as the SM prediction ($R_c\approx0.171$), one has
$R_b=0.2192\pm0.0018$ which is about 2$\sigma$ discrepancy with respect
to the expected value, $R_b=0.2156\pm0.0006$.

b) The second one, is the value of the left-right asymmetry
$A^{0e}_{LR}$
obtained by LEP measurements of the forward-backward asymmetry.
It corresponds to
$\sin^2\theta_W =0.2321\pm0.0005$\cite{lep} while SLD left-right
asymmetry measurement
implies $\sin^2\theta_W=0.2292\pm0.0010$\cite{sld}. This results are in
conflict with one another at about two standard deviations.
If confirmed, this could indicate new physics
coupled in a different way to the third generation.
For instance: 1)  extended gauge structures with
extra neutral bosons, like the $Z'$; 2) extra fermions like $t',b'$ or
even of heavy leptons as $E^-$;
3) non-standard Higgs particles, and  4) new heavy particles loop effects
like exotic leptons, quarks or supersymmetric particles.

It is possible that this will be an indication of new physics generating
a vertex correction to the $Z$ coupling or by new box contributions,
however the most exciting possibility is a new physics at tree
level.

All 3-3-1 models
have some of these requirements and no doubt they deserve to be study. In
particular model B has the flavor changing right-handed current in
Eq.~(\ref{rd2}). In fact, it has been
pointed out recently that the discrepancy between both measurements can be
reconciled if a new neutral gauge boson ${Z'}^0$ nearly
degenerate with the $Z^0$ do exist~\cite{ross}. This new neutral gauge
boson may also be responsible for the observed value of $R_b$.

Notice that according to Eqs. (\ref{ru2}) and (\ref{rd2}),
there are right-handed $u\leftrightarrow c\leftrightarrow t$ and
$d\leftrightarrow s \leftrightarrow t$ transitions mediated by the
$Z^0$ at tree level. In the charge $-1/3$ sector, the $K_L-K_S$
mass difference constrains only the matrix elements $(V^{D*}_R)_{3d}
(V^D_R)_{3s}$.
In order to determine the other elements of the matrix $V^D_R$ it is
necessary to study in detail $B$ decays.
A similar situation
occurs in the charge 2/3 sector.

On the other hand, all 3-3-1 models have both ${Z'}^0$ and extra fermions.
In particular there are heavy leptons in the models that we have considered
above~\cite{pt}. In these models it could
be necessary to take into account all new fields present in the
model.
Constraints coming from the neutral $K$ mass difference would not
necessarily imply a heavy $Z'$
since we can obtain consistence with the observed value of this mass
difference by choosing appropriately some of the matrix elements of $V^D_L$.
The later ones are different from the mixing angles appearing
in the observables $R_b$ and $A_{LR}$.
The couplings defined in Eq.~(\ref{nc}) for the case of the $Z'$ are
all flavor violating~\cite{mpp} and as we have extra mixing matrices
in these models it is possible that a global analysis of all data will show
compatibility among the low energy processes like $m_{K_L}-m_{K_S}$ mass
difference and the $Z$-peak observables. Exotic fermions can also give
contributions to the $Z\to b\bar b$ through loop effects.

\noindent {\em iv)} Some time ago it was pointed out that
since the left-handiness
of the $b$ quark has not been tested experimentally this
quark may decay through, in the extreme case,
purely right-handed couplings to the $c$ and $u$ quarks~\cite{gro,jr2}.
A test of the chirality of the $b$ quark
is the decay of polarized $\Lambda_b$ baryons.
These ideas were worked out in the context of
an $SU(2)_L\otimes SU(2)_R\otimes U(1)$ model. In such a model
the smallness of the $b$ to $c$ coupling is due not to the
value of the corresponding mixing angle but to the small value
of the right-handed Fermi constant $G_{FR}$, and the right-handed
$W_R$ boson must be light since \cite{rm}
\begin{equation}
\frac{G_{FR}}{G_{FL}}\simeq \frac{1}{\sqrt2}\left(g^2_R/M^2_{W_R}\right)
\left(g^2_L/M^2_{W_L}\right)\simeq V_{bc}\simeq 0.04.
\label{glr}
\end{equation}

In model B (Sec.\ref{subsec:mb}),
an intermediate situation
is realized. In Eq.~(\ref{wb}) the charged left-handed
currents are the usual ones. However, there are also right-handed currents
coupled to the $W^+$ boson with the same strength $G_F$ but it depends on some
of the right-handed couplings $V^U_R$ and $V^D_R$ appearing in Eq.~(\ref{wbr}).
Hence, the constraint in Eq.~(\ref{glr}) implies only that
$V^{U*}_{3cR}V^D_{3bR}<0.04$.

Notice that the left-handed coupling of the $b$ quark to the $c$ and $u$
quarks are the same of the Standard Model (See Eq.~\ref{wb}).
However, there are contributions to the semileptonic $b$-decays in which
a) a right-handed $b$-to-$c$($u$) current couples to a left-handed lepton
current (see Eq.~(\ref{cclb}) and (\ref{wbr}));
b) a left-handed $b$-to-$c$($u$)
current couples to a right-handed lepton current and, c) both currents are
right-handed. The cases b) and c) involve the heavy lepton sector:
$\bar \tau^c_{L}\gamma^\mu N_{1L}=-\bar N^c_{1R}\gamma^\mu \tau_{R}W_\mu^+$ or
$\bar T^c_L\gamma^\mu N_{2L}=-\bar N^c_{2R}\gamma^\mu T_RW_\mu^+$, and can be
suppressed if $N_i$ and $T$ are heavy.

Analyses of the $B_d^0-\bar B_c^0$ and $B^0_s-\bar B^0_s$ mixings
must be done in our context too. The dominant contributions in our model
come from two-$t$-quarks box diagrams as in the Standard Model. This involve
other matrix elements of $V^D_L$.
Hence, as we said before, in our models it would be necessary
to make a global analysis involving $Z$-pole observables, CP violation,
semileptonic $B$ decays and other processes in order to fit the several
parameters appearing in it.

\noindent {\em v)} These models predict new processes in which the
initial states have
the same electric charge as $ff\to W^-V^-$. This type of processes have only
recently begun to be studied~\cite{rs,pm}.
Also in some extensions of
these models, with spontaneous and/or explicit breaking of $L+B$
symmetry, it is possible to have
kaon decays with $\vert\Delta L\vert=2$, like
 $K^+\to\pi^-\mu^+\mu^+,\pi^-\mu^+e^+$
Similarly in $D$ and $B$ mesons decays.
Experimental
data imply $B(K^+\to \pi^-\mu^+\mu^+)<1.5\times10^{-4}$~\cite{rs}.
The process $e^-e^-\to W^-W^-$ which also could occur in some
extensions of the 3-3-1 models has been recently investigated in other
context~\cite{pm}.

In summary, none of these models is severely constrained at low energies.
For instance, in the leptonic
sector both of them are consistent with the existence
of three light neutrinos~\cite{pdg}.
Notice that in model B, the massless neutrinos (at tree level) $\nu_e,\nu_\mu$
do not mix with the heavy neutral fermions $N^0_i$ because of the symmetries
in Eqs.~(\ref{denovo}) and (\ref{dn2}). Mixing occurs only among
$\nu_\tau,\nu_T$
and $N^0_{iL}$. Thus it is not necessary to assume that
$H_i\langle\eta^0\rangle\gg\Gamma'\langle\sigma^0_1\rangle$ in Eq.~(\ref{ul}).
Neutrinos will get mass through radiative corrections
and some of their properties as the magnetic moments will be studied
elsewhere.

Another interesting
feature of this kind of models is that they include some extensions of
the Higgs sector which have been considered in the context of the
$SU(2)\otimes U(1)$ theory: more doublets, single and
doubly charged singlets, triplets, etc.

The supersymmetric version of the model of Ref~\cite{mpp} has been
considered in Ref.~\cite{ema}.

\acknowledgements

I would like to thank the
Con\-se\-lho Na\-cio\-nal de De\-sen\-vol\-vi\-men\-to Cien\-t\'\i
\-fi\-co e Tec\-no\-l\'o\-gi\-co (CNPq) for partial financial support.

\end{document}